\documentclass[aps, prb,nobalancelastpage,twocolumn,superscriptaddress,nolongbibliography]{revtex4-2}
\pdfoutput=1

\usepackage{color,amsthm,amsmath,amsxtra,amsfonts,dsfont,graphicx,bm,amssymb}

\usepackage[colorlinks=true,linkcolor=blue, citecolor=blue, urlcolor=blue, bookmarks]{hyperref}
\usepackage{centernot}

\usepackage[dvipsnames]{xcolor}

\newcommand{\T}{\mathcal{T}}

\newcommand{\be}{\begin{equation}}
\newcommand{\ee}{\end{equation}}

\newcommand{\beA}{\begin{align}}
\newcommand{\eeA}{\end{align}}

\newcommand{\bV}{\begin{bmatrix}}
\newcommand{\eV}{\end{bmatrix}}

\newcommand{\dd}{\mathrm{d}}
\newcommand{\de}{\partial}

\newcommand{\bea}{\begin{eqnarray}}
\newcommand{\eea}{\end{eqnarray}}
\newcommand{\bse}{\begin{subequations}}
\newcommand{\ese}{\end{subequations}}

\theoremstyle{plain}

\begin{document}
	
\title{Real-time spin-charge separation in one-dimensional Fermi gases \\
	from generalized hydrodynamics}

\begin{abstract}
We revisit early suggestions to observe  spin-charge separation (SCS) in cold-atom settings {in the time domain} by studying one-dimensional repulsive  Fermi gases in a harmonic  potential, where pulse perturbations are initially created at the center of the trap. We analyze the subsequent evolution using generalized hydrodynamics (GHD), which provides an exact description, at large space-time scales, for arbitrary temperature $T$, particle density, and interactions. At $T=0$  and vanishing magnetic field, we find that, after a nontrivial transient regime, spin and charge dynamically decouple up to perturbatively small corrections which we quantify. In this limit, our results can be understood based on a simple phase-space hydrodynamic picture. At finite temperature, we solve numerically the  GHD equations, showing  that for low $T>0$ effects of SCS survive and {characterize} explicitly  the value of $T$ for which the two distinguishable excitations melt. 
\end{abstract}

\author{Stefano Scopa}
\email{sscopa@sissa.it}
\affiliation{SISSA and INFN, via Bonomea 265, 34136 Trieste, Italy}
\author{Pasquale Calabrese}
\affiliation{SISSA and INFN, via Bonomea 265, 34136 Trieste, Italy}
\affiliation{International Centre for Theoretical Physics (ICTP), I-34151, Trieste, Italy}
\author{Lorenzo \surname{Piroli}}
\affiliation{Max-Planck-Institut f{\"{u}}r Quantenoptik, Hans-Kopfermann-Str. 1, 85748 Garching, Germany}

\maketitle


\section{Introduction}
The presence of peculiar physics in  one-dimensional ($1$D) quantum {systems} has been known for several decades, and is rooted in the dramatic effect of interactions compared with higher dimensions~\cite{korepin1997quantum,giamarchi2003quantum,gogolin2004bosonization,essler2005one}. 
The latter manifests itself in the collectivization and fractionalization of excitations, {with the most prominent and celebrated example being} the spin-charge separation  (SCS)~\cite{giamarchi2003quantum}{: In a 1D metal} a physical {charged spinful} particle splits into two different excitations for the spin and charge, respectively. 
{
SCS has been repeatedly probed since the late nineties in solid-state physics settings in the frequency domain~\cite{kim1996observation,segovia1999observation,tserkovnyak2003interference,auslaender2005spin,kim2006distinct,jompol2009probing,vianez2021observing}. 
In such experiments it is however not possible to access the time domain in which SCS presents more intuitive features.
To overcome these limitations, starting from the pioneering works by Recati {\it et al}. ~\cite{recati2003spin,recati2003fermi}, 
many real-time protocols have been proposed to probe SCS with cold atoms~\cite{recati2003spin,recati2003fermi,fuchs2005spin,kecke2005charge,liu2005signature,bohrdt2018angle,barfknecht2019dynamics,he2020emergence}.
Several experiments implemented these ideas in the past few 
years~\cite{boll2016spin,yang2018measurement,vijayan2020time}, 
but the field is still at an early stage}. 

{The aforementioned proposals~\cite{recati2003spin,recati2003fermi,fuchs2005spin,kecke2005charge,liu2005signature} are all}
based {on} an inhomogeneous Tomonaga-Luttinger-Liquid (TLL) 
approach~\cite{tomonaga1950remarks,luttinger1963exactly,haldane1981luttinger,giamarchi2003quantum,imambekov2012one} {for a}
confined $1$D {multicomponent gas} evolving after weak perturbations.
Then, SCS manifests itself in the real-time evolution of spin and charge profiles. Unfortunately, the TLL description applies only to very low temperatures and energies.
To understand the range of applicability of such approximation, direct simulations 
of the exact microscopic dynamics have been performed with tensor networks~\cite{kollath2005spin,kollath2006cold,kleine2008excitations,kleine2008spin}
and other numerical approaches~\cite{li2008collective,xianlong2008time,xianlong2010effects}.  
Still, tensor-network methods typically face a bottleneck due to the increasing computational cost at large times or finite temperatures~\cite{schollwock2011density}. 

\begin{figure}[t!]
	\includegraphics[width=0.45\textwidth]{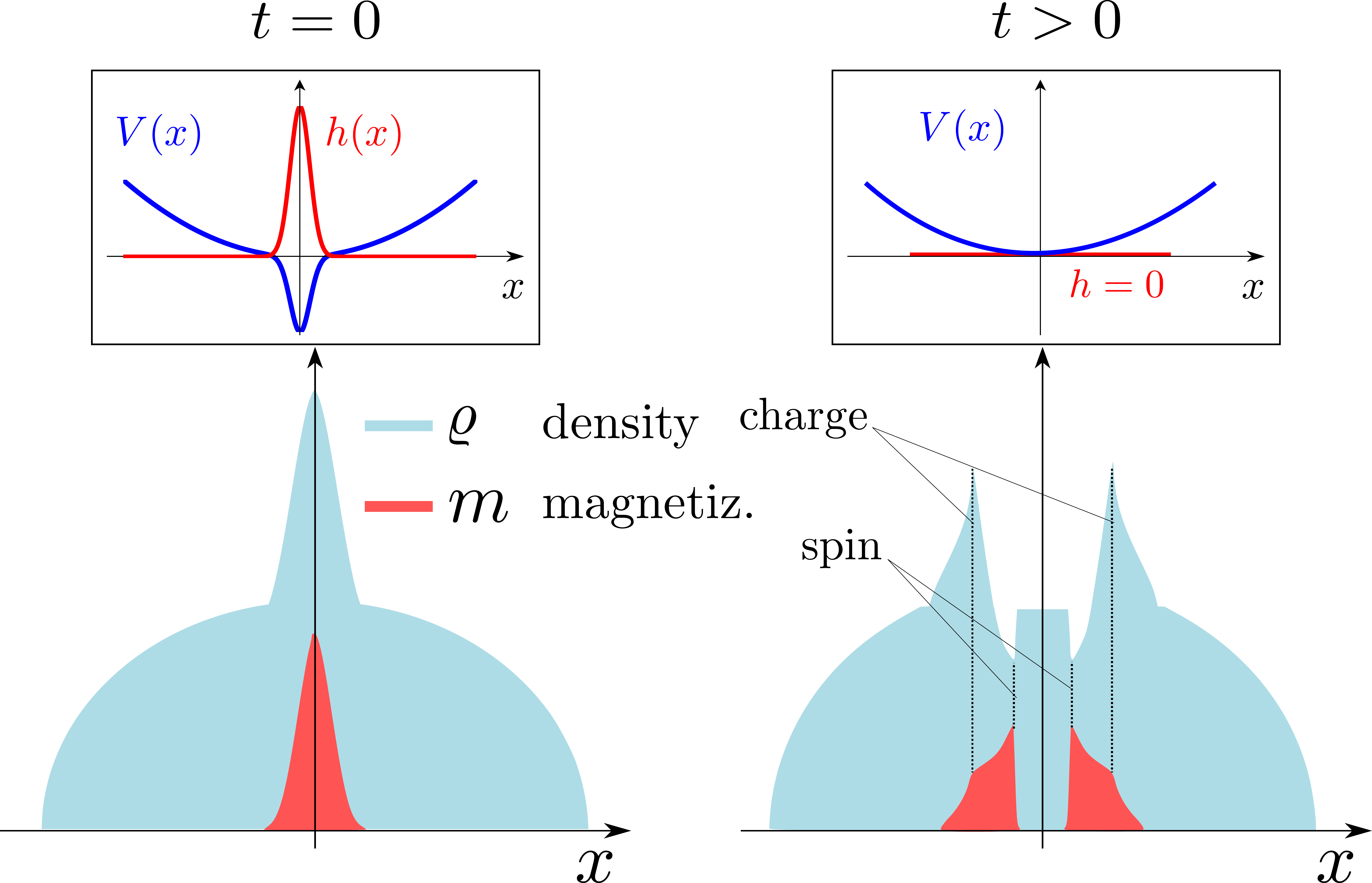}
	\caption{Pictorial representation of the protocol considered in this work. 
	A repulsive Fermi gas is confined in a harmonic trap, with superimposed potential wells {(top left)}. 
	At time $t=0$ the wells are removed, and the gas evolves in the harmonic trap {(top right)}.
	{The bottom plots are sketches of density and magnetization profiles.}}
	\label{fig:trap_quench}
\end{figure}

In this work, we revisit the problem by means of {the} recently developed generalized hydrodynamics (GHD)~\cite{bertini2016transport,castro-alvaredo2016emergent}, which gives access to the exact dynamics of integrable systems at hydrodynamic scales, for arbitrary interaction strength, temperature $T$, and particle density, see Refs.~\cite{alba2021generalized,bastianello2021hydrodynamics,denardis2021correlation} for reviews.
{For ultra-cold Bose gases subject to trap quenches, GHD proved to be powerful enough to provide quantitative predictions to perfectly match actual recent experimental results~\cite{schemmer2019,malvania2020generalized}, superseding conventional hydrodynamics \cite{doyon2017large}.
So far, GHD for multispecies quantum gases has been studied only in the idealized 
{\it bipartitioning protocol}~\cite{mestyan2019spin,bertini_transport_2019,wang2020emergent,nozawa2020generalized,nozawa2021generalized}, which is difficult 
to implement in real experiments.
Hence,}  
we study point wise repulsive Fermi gases in a harmonic potential, where finite-density spin and charge imbalances are initially created at the center of the trap, cf. Fig.~\ref{fig:trap_quench}. A similar setup has been investigated in Ref.~\cite{barfknecht2019dynamics} for a few-body Fermi system at strong interactions.
We predict SCS in the evolution of the profiles of charge and spin at low $T$, quantify finite-density corrections and explore the value  of $T$ for which it breaks down.\\
The rest of this work is organized as follows. In Sec.~\ref{sec:the_model} we introduce the model, which is exactly solvable by the Bethe Ansatz. We recall the aspects of the latter which will be relevant for us, including the thermodynamic description of the model, and the associated GHD equations. In Sec.~\ref{sec:pulse_dynamics} we consider an idealized setting, where the system is initially at zero temperature and infinitely large. This allows us to work out the main aspects of the problem without additional technical complications. The quantitative analysis of the real-time dynamics in the presence of trapping potential and finite temperature is finally presented in Sec.~\ref{sec:harmonic_confinement}, while our conclusions are reported in Sec.~\ref{sec:concluions}. The most technical aspects of our work are consigned to several appendixes.
\section{The model}\label{sec:the_model}
\subsection{The thermodynamic description}
We consider a $1$D gas of $N$ point wise interacting Fermions modeled by the Hamiltonian
\begin{align}
H=&\int_{-L / 2}^{L / 2} \mathrm{~d} x \sum_{\sigma=\uparrow, \downarrow} \hat{\Psi}_{\sigma}^{\dagger}(x)\left[-\partial_{x}^{2}+U(x,\sigma)\right] \hat{\Psi}_{\sigma}(x)\nonumber \\
 &+c \int_{-L / 2}^{L / 2} \mathrm{~d} x \ \hat{\Psi}_{\uparrow}^{\dagger}(x) \hat{\Psi}_{\downarrow}^{\dagger}(x) \hat{\Psi}_{\downarrow}(x) \hat{\Psi}_{\uparrow}(x)\,, \label{eq:hamiltonian}
\end{align}
where we set $\hbar=2m=1$. Here $\hat\Psi_{\sigma}(x)$, $\hat\Psi_{\sigma}^\dagger(x)$ are canonical fields associated with Fermions of spin up ($\uparrow$) and down ($\downarrow$), $c>0$ is the interaction strength, while $U(x,\sigma)=V(x)+\sigma h(x)-\mu$, where $V(x)$, $h(x)$ and $\mu$ are a longitudinal trap, the magnetic field and chemical potential, respectively. Note that the Hamiltonian preserves $N$, and the number $M$ of down spins.

For $V(x)\equiv 0$, $h(x)\equiv h$, the Hamiltonian is integrable~\cite{gaudin1967systeme,yang1967some} and can be solved using the {\it nested} Bethe Ansatz: The eigenfunctions can be written down exactly, and are labeled by two sets of numbers $\{k_j\}_{j=1}^N$,  $\{\lambda_j\}_{j=1}^M$, which are the quasimomenta (or \emph{rapidities}) of two distinct species of quasiparticles, {in some sense related to} the charge and spin degrees of freedom. In the thermodynamic limit, they allow for an intuitive description~\cite{takahashi2005thermodynamics}, which is reminiscent of that of noninteracting quantum gases. As a main difference, however, the spins can form $n$-quasiparticle bound states, and macrostates are then characterized by one charge rapidity distribution function $\rho_1(k)$,  and a set of functions for the spin $n$-quasiparticle bound states $\{\rho_{2,n}(\lambda)\}_{n=1}^\infty$, with $k,\lambda\in\mathbb{R}$.

Due to interactions, $\rho_1(k)$ and $\rho_{2,n}(\lambda)$ are not independent, but are related through the thermodynamic Bethe Equations (BE). These also involve  the distribution functions $\rho^h_1(k)$ and  $\rho^h_{2,n}(\lambda)$ for the {\it holes}, namely the vacant quasimomenta which can be occupied by the quasiparticles, see Appendix~\ref{sec:bethe_ansatz} for more detail. In particular, the BE reads
\begin{subequations}
\label{eq:BGT}
\begin{align}\label{eq:BGT1}
\rho_1(k) \!+\! \rho_1^h(k)&= \frac{1}{2\pi} + \sum_{n=1}^\infty [\phi_n\ast\rho_{2,n}](k)\,,\\
\label{eq: BGT2}
\!\!\!\rho_{2,n}(\lambda)\!+\!\rho_{2,n}^h(\lambda) &\!= \![\phi_n\!\ast\!\rho_1](\lambda) \!-\! \sum_{m=1}^\infty \![\Phi_{n,m}\!\ast \!\rho_{2,m}](\lambda)\,,
\end{align}
\end{subequations}
where 
\be
[f \ast g](\lambda)=\int_{-\infty}^{+\infty} d \nu f(\lambda-\nu) g(\nu)\,,
\ee
while 
\begin{align}
\phi_n(k)=& 2nc\pi^{-1}/[(nc)^2 + 4k^2]\,,\\
\Phi_{n,m}(k)=&(1-\delta_{n,m})\phi_{|n-m|}(k) + 2\phi_{|n-m|+2}(k)+\dots \nonumber\\
+& 2\phi_{n+m-2}(k) + \phi_{n+m}(k)\,.
\end{align}
The distribution functions uniquely fix the thermodynamic properties of the system. For instance, the density $\varrho=N/L$, and magnetization $m=(N/2-M)/L$ read 
\begin{align}
\varrho&=\int_{-\infty}^{+\infty} \dd k \rho_1(k),\ 
m=\frac{\varrho}{2}-\sum_{n=1}^\infty n\! \int_{-\infty}^{+\infty} \dd k \rho_{2,n}(k).\label{eq:def_magnetization}
\end{align}
These expressions have a simple interpretation, stating that thermodynamic quantities can be obtained as a weighted sum of single quasiparticle contributions, analogously to the noninteracting case.  {It is also custom to}  introduce the total distribution, 
\be
\rho^t_1(k)=\rho_1(k)+\rho^h_1(k)
\ee
and the Fermi factor, 
\be
n_1(k)=\rho_1(k)/\rho_1^t(k)
\ee
[$\rho^t_{2,n}(\lambda)$ $n_{2,n}(k)$  are defined analogously for the second species].
We stress that although all the nomenclature are taken from analogous free models, the interactions nontrivially dress all these functions in a way that is calculable only through
the solution of the BE.

\subsection{The GHD equations}
The above quasiparticle picture provides the foundation for GHD~\cite{bertini2016transport,castro-alvaredo2016emergent}. Within this framework, the system is described in terms of  Euler fluid cells, which are assumed to be well described by local quasistationary states~\cite{bertini2016determination} and whose properties are captured by  space-time-dependent rapidity distribution functions. Importantly, GHD applies also in the presence of external potentials~\cite{doyon2017anote}, provided that the typical lengths of variations of the local density are large compared with the microscopic scales~\footnote{See Refs.~\cite{doyon2017anote,doyon2017large} for detailed discussions on the validity of the hydrodynamic regime}. In particular, in this work we will consider a harmonic potential coupled to the density, i.e., $U(x,\sigma)=V(x)=\omega x^2$. The key result of GHD is a set of continuity equations for the rapidity distribution functions~\cite{bertini2016transport,castro-alvaredo2016emergent}, which in our case reads
\begin{subequations}\label{eq:GHD}
\begin{align}
\label{GHD1}
\de_t \rho_1(k) + \de_x [v_1(k)\rho_1(k)] =\left(\partial_{x} V\right) \partial_{k} \rho_1(k)\,,\\
\label{GHD2}
\de_t \rho_{2,n}(\lambda) + \de_x [v_{2,n}(\lambda) \rho_{2,n}(\lambda)] =\left(\partial_{x} V\right) \partial_{\lambda} \rho_{2,n}(\lambda)\,,
\end{align}
\end{subequations}
where we omitted explicit space-time dependence. Here $v_1(k)$ and $v_{2,n}(\lambda)$ are the quasiparticle \emph{dressed velocities}~\cite{bonnes2014light}. They are once again determined by the local rapidity distribution functions and can be computed solving the system
\begin{subequations}\label{eq:velocities}
	\begin{align}
v_1(k)\rho^t_1(k)&= \frac{k}{\pi} + \sum_{n=1}^\infty [\phi_n\ast v_{2,n}\rho_{2,n}](k)\,,\\
v_{2,n}(k)\rho_{2,n}(\lambda)&\!= \![\phi_n\!\ast\! v_1\rho_1](\lambda) \!-\! \sum_{m=1}^\infty \![\Phi_{n,m}\!\ast \!v_{2,m}\rho_{2,m}](\lambda)\,.
\end{align}
\end{subequations}
Equations~\eqref{eq:GHD} are exact in the limit where space-time scales of observations are sent to infinity, but have been shown to also provide extremely accurate results for relatively small times and sizes for appropriately ``smooth'' initial conditions~\cite{doyon2017large,bulchandani2017solvable,bulchandani2018bethe,lopez2021hydrodynamics}. They form the basis for the analysis reported in this work.
 
\section{Pulse-perturbation dynamics} \label{sec:pulse_dynamics}
Before analyzing SCS in the presence of a longitudinal potential, we consider an idealized scenario at zero temperature, where an infinite system is prepared in the ground state of $H$, with
\be
U(x,\sigma)=-\mu_0+ V(x)+\sigma h(x)\,,
\ee
$V(x)=-ae^{-x^2/w}$, and $h(x)=h_0e^{-x^2/w}$, creating an initial pulse perturbation~\cite{bettelheim2006orthogonality}. At $t=0$, the potential and field are switched off, $V\equiv h\equiv 0$, letting the system evolve freely. Here, the main features of SCS can be deduced analytically, as we now discuss.

\begin{figure}
	\centering
	\includegraphics[width=0.5\textwidth]{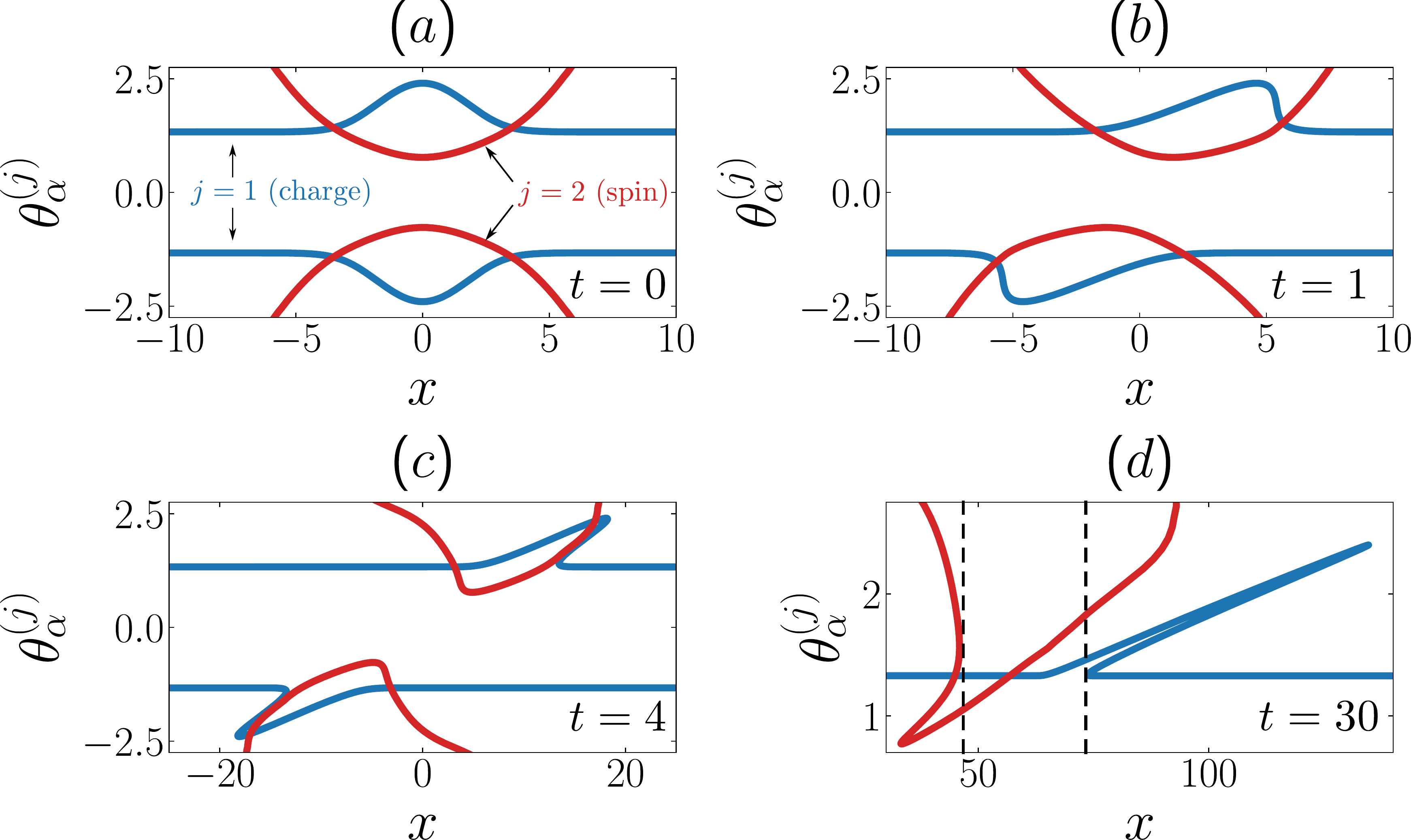}\\
	\caption{$(a)$: Fermi contours for the ground state in the presence of Gaussian potentials, cf. the main text. By definition, the Fermi factors $n_j(k)$ are $1$ for all points $(x,k)$ inside the contour, and $0$ outside. Far from the origin, $h\to 0$, and the Fermi rapidities $\theta^{(2)}_\alpha$ diverge. $(b)$-$(d)$: Snapshots of the Fermi contours at times $t=1,4, 30$.  In $(d)$, dashed lines correspond to the position $x_{j}(t)$ of the max. (min.) connected Fermi sea with $j=1$ ($j=2$), associated with the peaks in the charge and magnetization profiles. In this figure, $c=1$, $\mu=1.5$, $a=2.25$, $h_0=2$ and $w=5$.}
	\label{fig:contour_evolution}
\end{figure}

First, we use a local-density approximation (LDA)~\cite{cazalilla2011one} to obtain the initial conditions of the GHD equations. Namely, at $t=0$, we associate with a fluid cell at position $x$ the ground-state of the homogeneous Hamiltonian with $\mu=\mu_0-V(x)$ and $h=h(x)$,  whose distribution functions can be computed by Bethe Ansatz, cf. Appendixes~\ref{sec:bethe_ansatz} and~\ref{sec:zero_temperature_limit}. By construction, each local state has initially zero entropy and, since the latter is conserved by the GHD equations~\cite{doyon2017anote}, fluid cells remain in zero-entropy states at all times. This is important, because for integrable models the latter are known to be ``split Fermi seas''~\cite{fokkema2014split,eliens2016general,vlijm2016correlation}, leading to very simple dynamics~\cite{doyon2017large}. In our case, these are states characterized by no spin bound states, $n_{2,n}(k)\equiv 0$, and Fermi factors $n_1(k)$ and $n_2(k)=n_{2,1}(k)$ of the form
 \be\label{zero-entropy}
n_j(k)=\left\{\begin{array}{ll}1 & \text { if } k\in[\theta^{(j)}_{1}, \theta^{(j)}_{2}] \cup \cdots \cup[\theta^{(j)}_{2 q-1}, \theta^{(j)}_{2 q}]\,, \\ 0 & \text { otherwise. }\end{array}\right.
\ee
Thus, any fluid cell is fully described by the \emph{Fermi rapidities} $\{\theta^{(j)}_{\alpha}\}$, representing the edges of the split Fermi seas, and all the information on the system is encoded into a pair of time-dependent \emph{Fermi contours} $\Gamma^{(j)}_t$, keeping track of the Fermi rapidities at each point in space~\cite{doyon2017large}. For $t=0$, the Fermi contours are obtained from LDA, taking the form displayed in Fig.~\ref{fig:contour_evolution}$(a)$; for each species there is a single pair $(\theta^{(j)}_{1}(x), \theta^{(j)}_{2}(x))$ for all positions $x$, with $ |\theta^{(2)}_{1}(x)|\propto x^2$, for large $x$, while $|\theta^{(j)}_{1}(x)|$ approaches a constant. The dynamics of the Fermi contours follow from the zero-temperature limit of Eq.~\eqref{eq:GHD}. In the absence of potentials, it reduces to the wave equation~\cite{doyon2017large} 
\be
\partial_{t} \theta_{\alpha}^{(j)}+v_j(\theta_{\alpha}^{(j)}) \partial_{x} \theta_{\alpha}^{(j)}=0\,,
\ee 
where $v_1(\theta)$, $v_2(\theta)=v_{2,1}(\theta)$ are the dressed velocities; see Appendix~\ref{sec:GHD_equations} for more detail. Importantly, they are positive (negative) in the upper (lower) half of the $x-\theta^{(j)}$ plane. Furthermore, for both species they are increasing functions of the Fermi points $\theta^{(j)}$, with the velocity of the second species being always smaller than the first one.

\begin{figure}
	\includegraphics[width=0.4\textwidth]{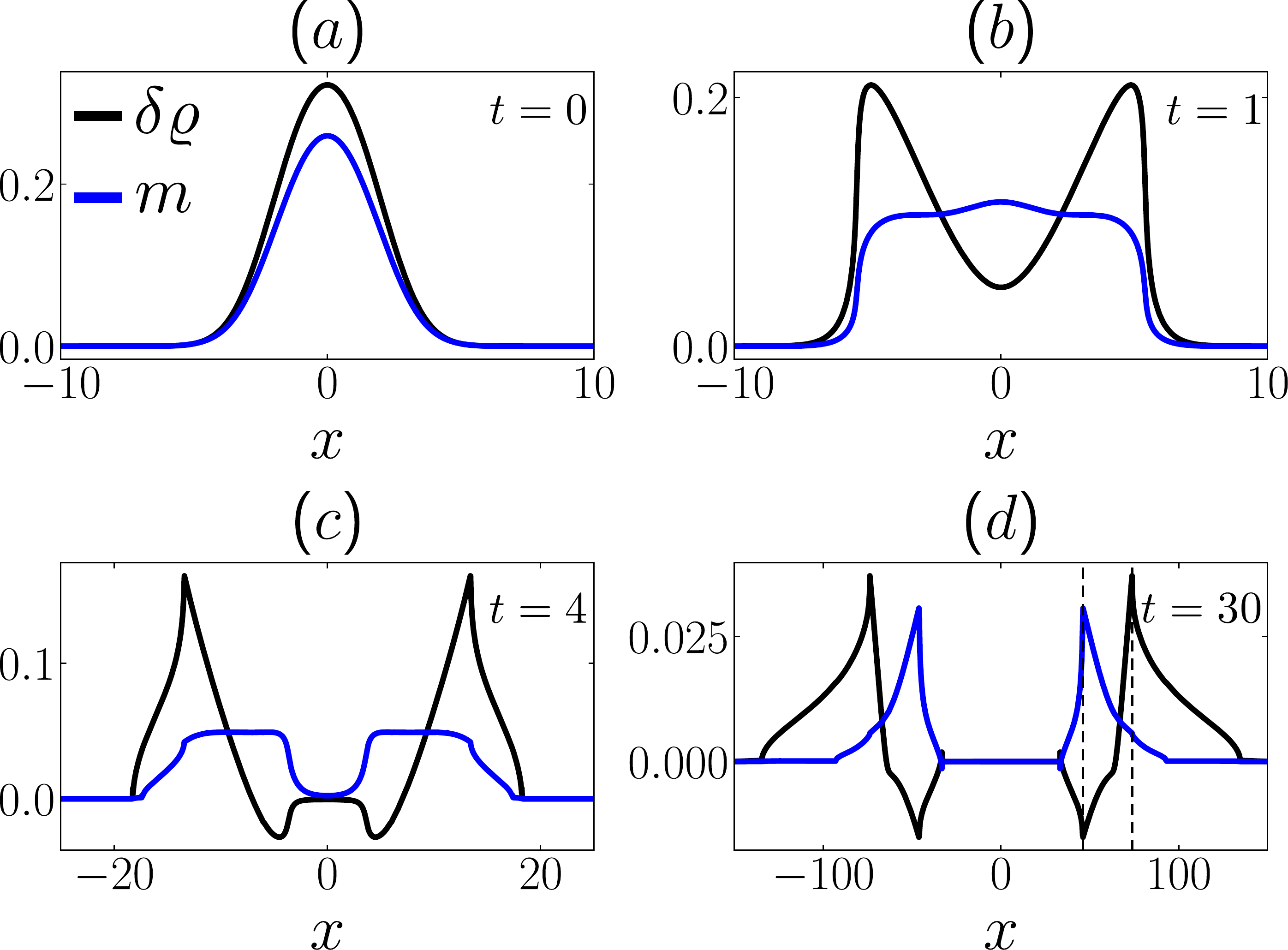}
	\caption{Snapshots of the magnetization and excess density $\delta\varrho=\varrho-\varrho(\infty)$ for $t=0,1,4,30$, after release from Gaussian potentials. In $(d)$, dashed lines are at the same position of those shown in Fig.~\ref{fig:contour_evolution}$(d)$ and correspond to the peaks of the profiles. The two propagate at different velocities and are independent, up to small corrections quantified in the main text. }
	\label{fig:profile_evolution}
\end{figure}

Based on these considerations, we can understand qualitatively the dynamics of the contours, cf. Fig.~\ref{fig:contour_evolution}. Focusing on the upper half of the $x-\theta^{(j)}$plane, the charge and spin contours move on the right at different velocities. Although the evolution is initially nontrivial, since the extremum of the charge (spin) moves with largest (smallest) speed, we can predict that the bumps initially located near the origin will eventually overturn. This means that, in some spatial region, both Fermi seas split into disconnected components, going from having two to four Fermi rapidities, cf. Fig.~\ref{fig:contour_evolution}$(c)$. This reflects in the formation of well-defined peaks in the spin and charge profiles, which are obtained using Eq.~\eqref{eq:def_magnetization}, cf. Fig.~\ref{fig:profile_evolution}. We claim that, at large times, these peaks propagate independently, up to perturbatively small corrections, signaling a dynamical separation of spin and charge.

In order to show this, we first analyze the local macrostate at the main peak in the profile of the charge, $x_1(t)$, which is the largest position for which  the Fermi sea of the first species is not split, cf. Fig.~\ref{fig:contour_evolution}$(d)$. At $x_1(t)$, we have two Fermi points $A_\pm(t)=\theta^{(1)}_{1,2}(x_1(t))$ for the charge, and two for the spins  $B_{\pm}(t)=\theta^{(2)}_{1,2}(x_1(t))$. In this case, the rapidity distribution functions are given by
\begin{subequations}
\label{eq:bgt}
\begin{align}
\rho_1(k)&=\frac{1}{2\pi}+ [\phi_1\ast\rho_{2}]_{B_-}^{B_+}(k),\label{eq:bgt1}\\
\rho_{2,1}(\lambda)&=[\phi_1\ast\rho_1]_{A_-}^{A_+}(\lambda) - [\phi_2\ast\rho_{2,1}]_{B_-}^{B_+}(\lambda),\label{eq:bgt2}
\end{align}
\end{subequations}
where 
\be
[f\ast g]_{\alpha}^\beta(x)=\int_{\alpha}^\beta \dd yf(y-x)g(y)\,.
\ee
For large $t$, we have  $x_1(t)\sim t$, and since the velocity of the second species is always smaller that the first one, we also have $|B_\pm(x_1(t))|\sim x_1(t)^2 \sim t^2$. Plugging this information into Eq.~\eqref{eq:bgt}, it is possible to show analytically that the magnetization $m$ at $x_1(t)$ is exponentially small in $t$. This follows from an asymptotic analysis of the solution of Eq.~\eqref{eq:bgt} in the limit of large $|B_{\pm}|$, and Eq.~\eqref{eq:def_magnetization}. Since this is rather technical, we report it in Appendix~\ref{sec:perturbative_scs}. As a consequence, the charge pulse does not generate a variation in the background magnetization, thus propagating independently from the latter.

A more careful analysis is needed for the macrostate at the main peak in the profile of the magnetization,  $x_2(t)$. From Fig.~\ref{fig:profile_evolution}, {the spin} propagation is not completely decoupled from the charge, as we observe a visible perturbation in the profile of the charge at $x_2(t)$. This is a manifestation that $\rho_1(k)$ and $\rho_{2,1}(\lambda)$ are coupled through the BE~\eqref{eq:bgt}. However, denoting by $\Delta \varrho$ and $\Delta m$ the peak values of charge and magnetization at $x_2(t)$, from a careful analysis of the BE~\eqref{eq:bgt} it can be shown that 
\be\label{eq:final_result_perturbative}
|\Delta \varrho|\sim |\Delta m|^2\,.
\ee
This is an important result: It states that the charge pulse generated by the propagating spin wave is perturbatively small and at the leading order in the spin-wave amplitude we observe the previously announced decoupling of the spin and charge degrees of freedom.
The derivation of Eq.~\eqref{eq:final_result_perturbative} is quite technical, and is therefore reported in Appendix~\ref{sec:perturbative_scs}. This prediction, together with its interpretation using  the zero-entropy GHD framework, is the first main result of this work.

We stress that the formalism above is similar to the phase-space hydrodynamics used to study pulse propagation in free Fermi gases~\cite{bettelheim2006orthogonality,bettelheim2008quantum,bettelheim2012quantum,protopopov2013dynamics,kulkarni2018}, nonlinear Luttinger liquids~\cite{dissipationless2014protopov}, (spin-)Calogero~\cite{abanov2005quantum,abanov2009integrable,kulkarni2009nonlinear,kulkarni2011cold} and Lieb-Liniger models~\cite{damski2006shock,peotta2014quantum,sarishvili2016pulse}. As a main difference, however, here the velocities are dressed by the interactions, which are also responsible for the nontrivial interplay between spin and charge dominating the dynamics at short times. We also note that, while the GHD equations are exact, LDA is not. However, while corrections are expected, they can be made arbitrarily small by reducing the spatial density variations in the initial conditions~\cite{bettelheim2011universal, doyon2017anote}.
 
\section{Harmonic confinement and nonzero temperatures} \label{sec:harmonic_confinement}
In this section, we finally turn to the case of a confining harmonic potential and $T>0$.

The effect of the trap can be easily taken into account, modifying our previous solution in two ways. First, the potential changes the initial condition of the GHD equations: From LDA, we now obtain that the gas is confined in a finite interval $I_0=[-x_{0},x_0]$. Second, it modifies the dynamics, yielding a nonvanishing driving term in Eq.~\eqref{eq:GHD}. Based on these considerations, we can predict that SCS effects survive provided that the trap is {shallow enough}. Namely,  the dynamics with and without the trap remains qualitatively similar as long as the position of the propagating pulses, $|x_{1,2}(t)|$, is much smaller than the spatial extent of the trap, $|x_{1,2}(t)|\ll x_0$. 

\begin{figure}
	\includegraphics[width=0.5\textwidth]{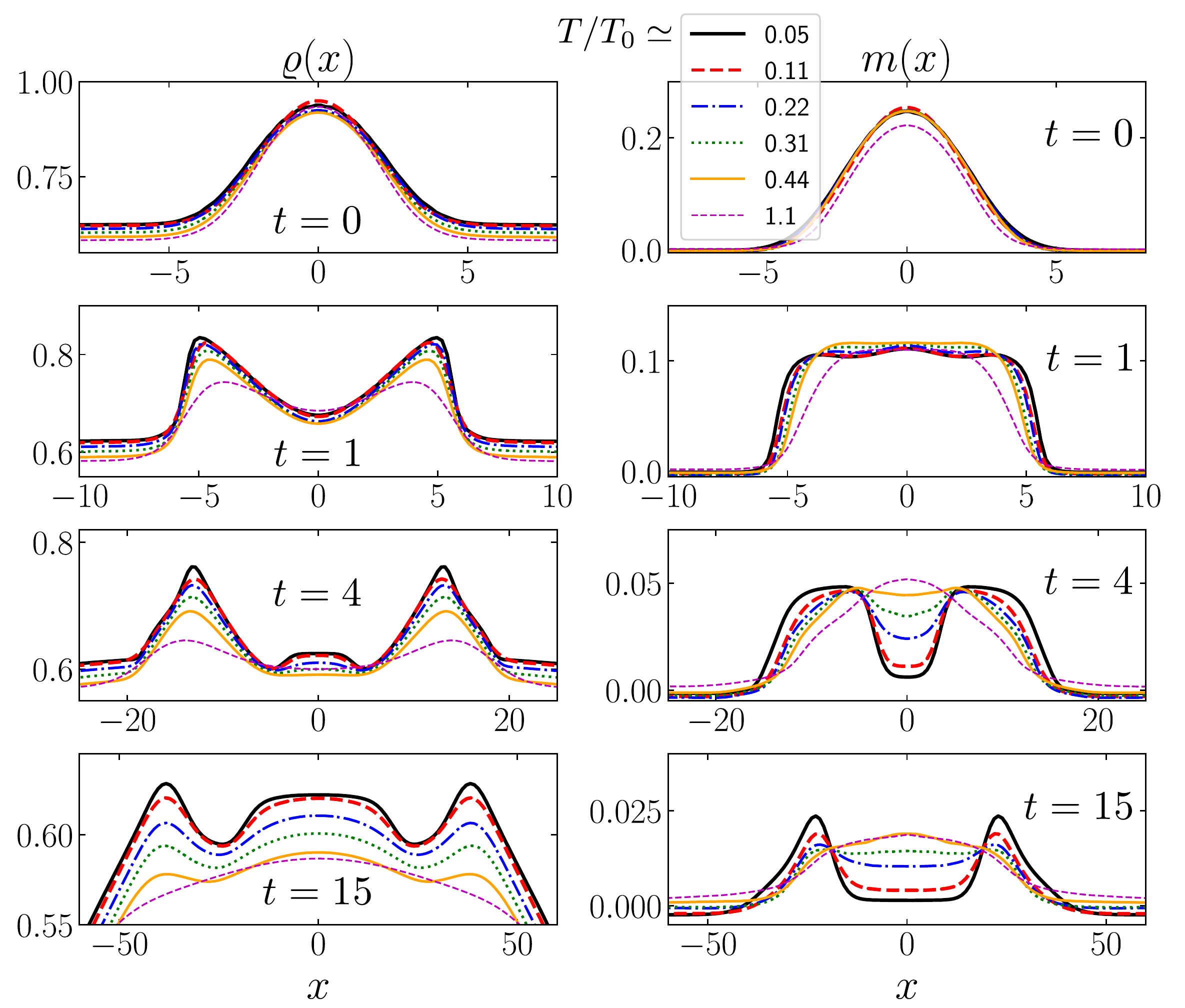}
	\caption{Snapshots of the magnetization and density for $t=0,1,4,15$, after release into a harmonic trap, for increasing temperatures $T$. Here $T_0$ is defined from the density and Fermi rapidities of the background at the center of the trap, where $k_0=1.33$, $\varrho_0=0.623$. For the harmonic trap we chose $\omega=0.01$ while the other parameters are set as in Fig.~\ref{fig:contour_evolution}. {These profiles clearly show SCS as discussed in the text.}}
	\label{fig:profile_zero_temperature}
\end{figure}

A nonvanishing temperature, on the other hand, calls for a more careful analysis. Let us consider first an infinite system, initially prepared at temperature $T$. For $T=0$, the dynamics of the pulses propagating away from the origin can be understood in terms of excitations over a constant background, which is the local equilibrium macrostate at $x=\pm\infty$, with particle density $\varrho_0$ and Fermi rapidities (for the first species) $\pm k_0$. As the density of excitations becomes small at large times, one enters the regime of validity of the TLL theory~\footnote{See also~\cite{bertini2018universal,bertini2018low}, where the emergence of a Luttinger-Liquid description from the GHD equations has been established rigorously in bipartitioning protocols.}, and the peaks of spin and charge propagating  at different velocities can be understood in terms of decoupled Luttinger liquids. Thus SCS effects are expected to survive at least up to temperature scales for which the standard TLL description applies. In this case, it is known that the relevant temperature scale is $T_0=[k_0^2/\gamma]^{-1}$ with $\gamma=c/\varrho_0$, above which the gas enters the incoherent Luttinger-Liquid regime~\cite{fiete2007colloquium}. Now, for a finite system in a sufficiently long trap,  the density and the Fermi rapidities are approximately constant over the spatial region where the initial dynamics takes place, i.e., $\rho(x)\simeq \rho_0$, $k_0(x)\simeq k_0$ yielding the temperature scale $T_0$ of  the problem. 

We stress that these estimates for the temperatures at which SCS effects survive are difficult to justify rigorously in the present nonequilibrium setting and should be corroborated by a quantitative analysis of the profiles. This  constitutes our second main result. We have solved the GHD equations~\eqref{eq:GHD} in the presence of a long trap and at different temperatures. An example of our data is reported in Fig.~\ref{fig:profile_zero_temperature}. In general, we find that SCS effects remain clearly visible for $T\ll T_0$. Already for very low temperatures, however, the profiles are smeared out compared with $T=0$, and the corresponding nonanalytic points disappear. Our calculations confirm that the peaks of the two species remain visible up to $T\simeq T_0$. Above this threshold, {the spin} pulses melt while the charge ones remain distinguishable, consistently with what could be expected from the spin-incoherent Luttinger-Liquid theory~\cite{fiete2007colloquium}. Increasing further the temperature, the signals in the charge profiles will become eventually
invisible. Finally, we found that the effect of the harmonic trap is weak for the observed time scales, provided that its length is large compared with the region where the pulses propagate.

\section{Conclusions}\label{sec:concluions} 
We have studied SCS in confined $1$D Fermi gases where pulse perturbations are initially created at the center of the trap. Our results are based on GHD, which allows us to obtain an exact description for arbitrary temperature, particle density, and interactions. While we have focused on one particular protocol, more general settings can be analyzed using the same approach, including nonzero magnetic fields, trapping potentials or geometric quenches \cite{next-publ}. 
{As a follow-up, it would also be important to quantify the effect of quantum fluctuations on top of the GHD solution, adapting recent approaches for Bose gases and spin-chains \cite{Fagotti2017high-ord-hydro,Ruggiero2019ConfBreathing,Ruggiero2019QGHD,Collura2020DomainWallMelting,Scopa2021exact}}.
Overall, our work shows how GHD is capable of predicting SCS effects in interacting multicomponent quantum gases in versatile experimentally relevant situations.

\section*{Acknowledgments} 
P.C. and S.S. acknowledge support from ERC under Consolidator Grant No. 771536 (NEMO). L.P. acknowledges support by the DFG (German Research Foundation) under Germany’s Excellence Strategy --  EXC-2111 -- 390814868.
We are very grateful to Leonardo Fallani and Jacopo Catani for discussions about possible experimental implementations of our results. \\[4pt]
\noindent
{\bf Note added.}~ After the publication of this paper, we discovered a typo appearing in Eq.~\eqref{GHD1} of the main text and in Eqs.~\eqref{eq:appendix_GHD2},\eqref{eq:ghd2}, \eqref{eq:phase-space} of the appendices. We fixed these typos in this updated version. All the results in the published version remain quantitatively correct .


\appendix 

\section{The Bethe Ansatz solution}
\label{sec:bethe_ansatz}

In this appendix we provide more detail on the Bethe Ansatz solution of the model. In the absence of external potentials  and for constant magnetic field, the Hamiltonian~\eqref{eq:hamiltonian} is integrable and defines  the Yang-Gaudin model~\cite{yang1967some,gaudin1967systeme}. As a consequence, there exists an infinite number of local conserved operators $\{Q_{n}\}_{n=1}^\infty$ (usually called charges) commuting with the Hamiltonian, which can be solved via the (nested) Bethe Ansatz~\cite{takahashi2005thermodynamics,essler2005one}. Here we  report  the aspects of the solution which are directly relevant for our work, while we refer to the literature for a more systematic treatment~\cite{korepin1997quantum,takahashi2005thermodynamics,essler2005one}. 

The eigenfunctions  of~\eqref{eq:hamiltonian} can be labeled by two sets of rapidities $\{k_j\}_{j=1}^N$,  $\{\lambda_j\}_{j=1}^M$, associated with the charge and spin degrees of freedom. Due to periodic boundary conditions, the rapidities are quantized, and must satisfy the following algebraic equations:
\begin{align}\label{eq:appendix_BE}
e^{ik_j L}&= \prod_{\alpha=1}^M \frac{k_j -\lambda_\alpha+ ic/2}{k_j -\lambda_\alpha-ic/2}\,,\\
\prod_{j=1}^N \frac{\lambda_\alpha-k_j+ic/2}{\lambda_\alpha-k_j -ic/2}& = \prod_{\beta\neq \alpha,\atop \beta=1}^M \frac{\lambda_\alpha-\lambda_\beta+ic}{\lambda_\alpha-\lambda_\beta-ic}\,,
\end{align}
which are known as BE. The energy of a given eigenstate then simply reads
\be
E[\{k_j\}_{j=1}^N,\{\lambda_{\alpha}\}_{\alpha=1}^M]=-\mu N+2h M+\sum_{j=1}^N e(k_j)\,,
\ee
where $e(k)=k^2$.

For large values of the system size $L$, the solutions to Eq.~\eqref{eq:appendix_BE} satisfy the string hypothesis~\cite{takahashi2005thermodynamics}, according to which $k_j$ are real, while the rapidities $\lambda_\alpha$ form patterns in the complex plane called strings. 
An $n$-string consists of $n$ rapidities distributed symmetrically around the real axis, with the $j$-th rapidity in the string being
\be
\lambda_{\alpha,j}= \lambda_\alpha^n + i(n+1-2j)c/2, \quad j=1,\dots, n\,.
\ee
$\lambda_\alpha^n\in\mathbb{R}$ is known as the string center. The string hypothesis is expected to be true  up to exponentially small corrections in the system  size. In the thermodynamic limit $ L\to \infty$,  $N/L={\rm cst}$, $M/L={\rm cst}$, the spectrum of the model becomes densely populated and we can adopt a description in terms of the rapidity distribution functions formally defined by
\be
\rho_1(k_j)\sim \frac{1}{L (k_{j+1}-k_j)}\,,\qquad \rho_{2,n}(\lambda_{\alpha}^n)\sim\frac{1}{L (\lambda_{\alpha+1}^n -\lambda_\alpha^n)}\,.
\ee
Due to interactions, $\rho_1(k)$ and $\rho_{2,n}(\lambda)$ are not independent. Their relation can be found by taking the thermodynamic limit of the BE~\eqref{eq:appendix_BE}, yielding
\begin{subequations}
	\label{eq:_appendix_BGT}
	\begin{align}\label{eq:_appendix_BGT1}
		\rho_1(k) \!+\! \rho_1^h(k)&= \frac{1}{2\pi} + \sum_{n=1}^\infty [\phi_n\ast\rho_{2,n}](k)\,,\\
		\label{eq:_appendix_BGT2}
		\!\!\!\rho_{2,n}(\lambda)\!+\!\rho_{2,n}^h(\lambda) &\!= \![\phi_n\!\ast\!\rho_1](\lambda) \!-\! \sum_{m=1}^\infty \![\Phi_{n,m}\!\ast \!\rho_{2,m}](\lambda)\,,
	\end{align}
\end{subequations}
where 
\begin{align}
[f \ast g](\lambda)&=\int_{-\infty}^{+\infty} d \nu f(\lambda-\nu) g(\nu)\,,\\
\phi_n(k)&=\frac{1}{\pi}\frac{2nc}{(nc)^2 + 4k^2}\,,\\
\Phi_{n,m}(k)&=(1-\delta_{n,m})\phi_{|n-m|}(k) + 2\phi_{|n-m|+2}(k)+\dots\nonumber\\
 +& 2\phi_{n+m-2}(k) + \phi_{n+m}(k)\,.
\end{align}
Note that the thermodynamic BE~\eqref{eq:_appendix_BGT} also involve  the distribution functions $\rho^h_1(k)$ and  $\rho^h_{2,n}(\lambda)$ for the holes, 
i.e., the vacancies in the space of rapidities which can be occupied by the  quasiparticles. In the noninteracting case, the distribution functions for the holes and quasiparticles are trivially related. However, this is not the case in the presence of interactions. For this reason, it is convenient to also introduce the total distributions
\be
\rho^t_1(k)=\rho_1(k)+\rho^h_1(k),\ \rho^t_{2,n}(k)=\rho_{2,n}(k)\!+\!\rho^h_{2,n}(k),
\ee
the Fermi factors
\be\label{eq:appendix_fermi_factors}
n_1(k)=\frac{\rho_1(k)}{\rho_1^t(k)}\,,\qquad n_{2,n}(k)=\frac{\rho_{2,n}(k)}{\rho_{2,n}^t(k)}\,,
\ee
and the  closely related functions
\be
\eta_1(k)=\frac{\rho^h_1(k)}{\rho_1(k)}\,,\qquad \eta_{2,n}(k)=\frac{\rho^h_{2,n}(k)}{\rho_{2,n}(k)}\,.
\ee

\begin{figure*}
	\centering
	(a) \hspace{3cm} (b)\\
	\includegraphics[width=0.75\textwidth]{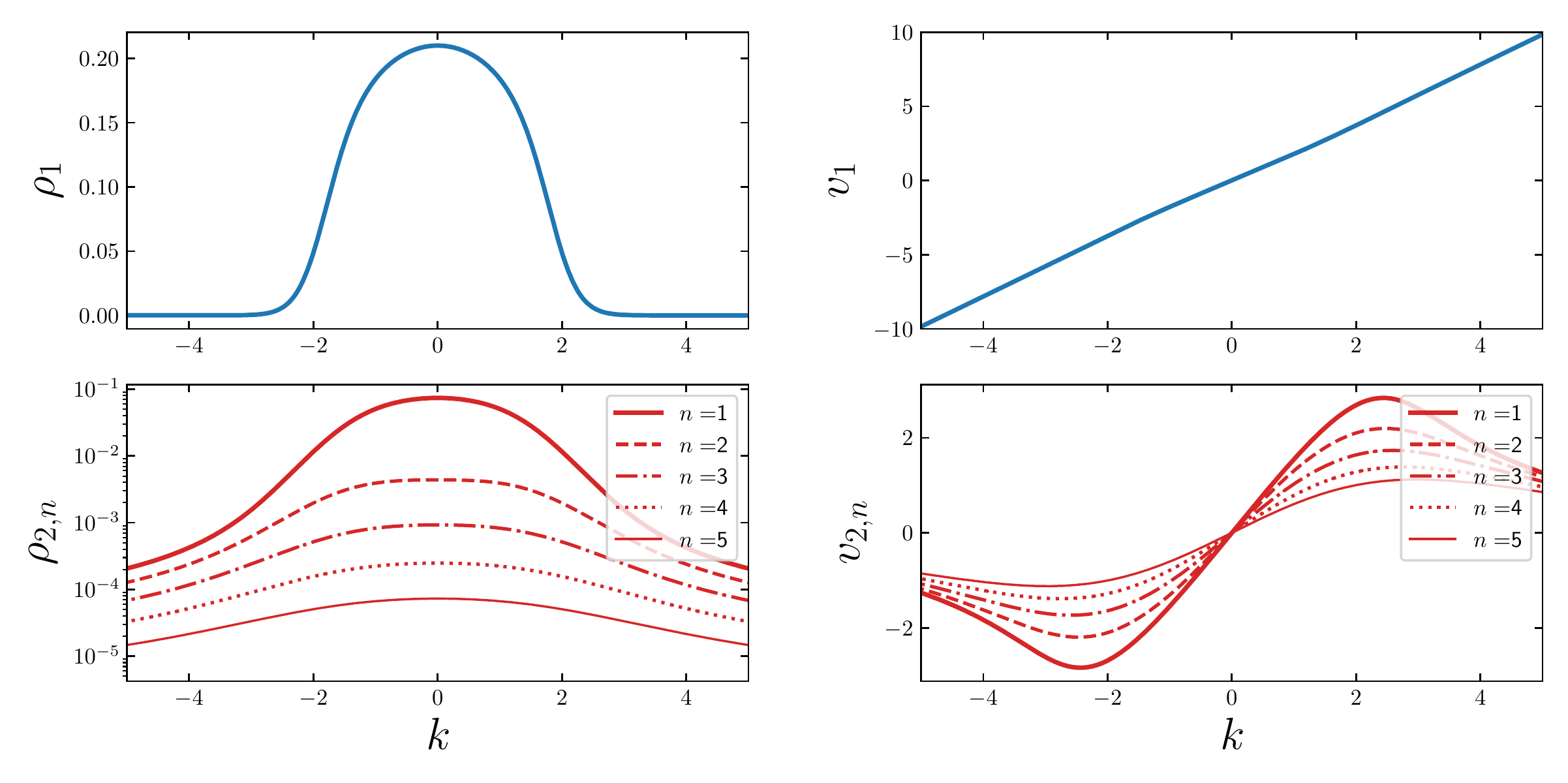}
	\caption{(a) Rapidity distribution functions and (b) effective velocities as a function of the quasimomentum $k$, obtained as a numerical solution of Eqs.~\eqref{eq:_appendix_BGT} (panel a) and Eqs.~\eqref{eq:velocities_TBA} (panel b). We have set $c=1$, $\mu=2$, $h=0.5$ and $T=1$.}
	\label{fig:root-dens}
\end{figure*}

The BE~\eqref{eq:_appendix_BGT} do not uniquely specify a single  set of rapidity distribution functions, and an additional set of integral equations is  typically needed in order to determine them. An important example is that of thermal states, which are characterized by a given temperature $T$. In this case, the additional set of integral equations can be obtained within the  thermodynamic  Bethe Ansatz formalism~\cite{takahashi2005thermodynamics} and read
\begin{subequations}
\label{eq:TBA}
\begin{align}
\ln\eta_1(k)&= \frac{k^2-\mu-h}{T} -\sum_{n=1}^\infty[\phi_n\ast \ln(1+\eta_{2,n}^{-1})](k)\,,\label{eq:TBA1}\\
\ln\eta_{2,n}(k)=& \frac{2nh}{T} -[\phi_n\ast\ln(1+\eta_{1}^{-1})](k) \nonumber\\
+& \sum_{m=1}^\infty [\Phi_{n,m}\ast \ln(1+\eta_{2,n}^{-1})](k)\,.\label{eq:TBA2}
\end{align}
\end{subequations}
Eqs.~\eqref{eq:TBA} and \eqref{eq:_appendix_BGT} can be efficiently solved numerically with iterative methods, yielding the desired rapidity  distribution functions, cf. Fig.~\ref{fig:root-dens}(a) for an example. Once they are known, one can immediately compute several thermodynamic quantities of the model. 
For instance, the particle and magnetization  densities are given by
\begin{align}
\varrho&=\frac{N}{L}=\int_{-\infty}^\infty \dd k \ \rho_1(k)\,,\\
m&=\frac{N-2M}{2L}=\frac{\varrho}{2}-\sum_{n=1}^\infty n \int_{-\infty}^\infty \dd k \ \rho_{2,n}(k)\,.
\end{align}
More generally, a similar expression exists for the expectation value $\mathcal{Q}$ of any conserved charge, i.e.,
\be\label{eq:charge}
q=\frac{\cal Q}{L}=\int_{-\infty}^\infty \dd k \ q_1(k) \rho_1(k) + \sum_{n=1}^\infty \int_{-\infty}^\infty \dd \lambda  \ q_{2,n}(\lambda) \rho_{2,n}(\lambda)\,,
\ee
where $q_1(k)$, $q_{2,n}(\lambda)$ are known functions. These expressions have a simple interpretation, stating that thermodynamic quantities can be obtained as a weighted sum of single quasiparticle contributions, analogously to the noninteracting case. Finally, together with the expectation values of the charges,  rapidity distribution  functions also allow us to compute the dressed velocities of the quasiparticles in an arbitrary macrostate. In particular, they can be obtained as the solution to the equations~\cite{bonnes2014light}
\begin{subequations}
\label{eq:velocities_TBA}
\begin{align}
v_1(k) \rho^{t}_1(k)&= \frac{k}{\pi} + \sum_{n=1}^\infty[\phi_n\ast v_{2,n} \rho_{2,n}](k)\,,\label{eq:velocities_TBA_1}\\
	v_{2,n}(k) \rho^{t}_{2,n}(k)&=[\phi_n  \ast v_1 \rho_1](k)  -\sum_{m=1}^\infty [\Phi_{n,m} \ast v_{2,m} \rho_{2,m}](k)\,.\label{eq:velocities_TBA_2}
\end{align}
\end{subequations}
As an example, we report in Fig.~\ref{fig:root-dens}(b) their numerical solution for a particular thermal state. 

\section{Zero-temperature limit}
\label{sec:zero_temperature_limit}

In the zero-temperature limit, the thermodynamic description of the model becomes particularly simple. Here we sketch the main formulas, referring once again to the literature for more details~\cite{takahashi2005thermodynamics}. 

\begin{figure*}
	\centering
(a) \hspace{3.5cm} (b) \hspace{5cm} (c)\\
	\includegraphics[width=0.45\textwidth]{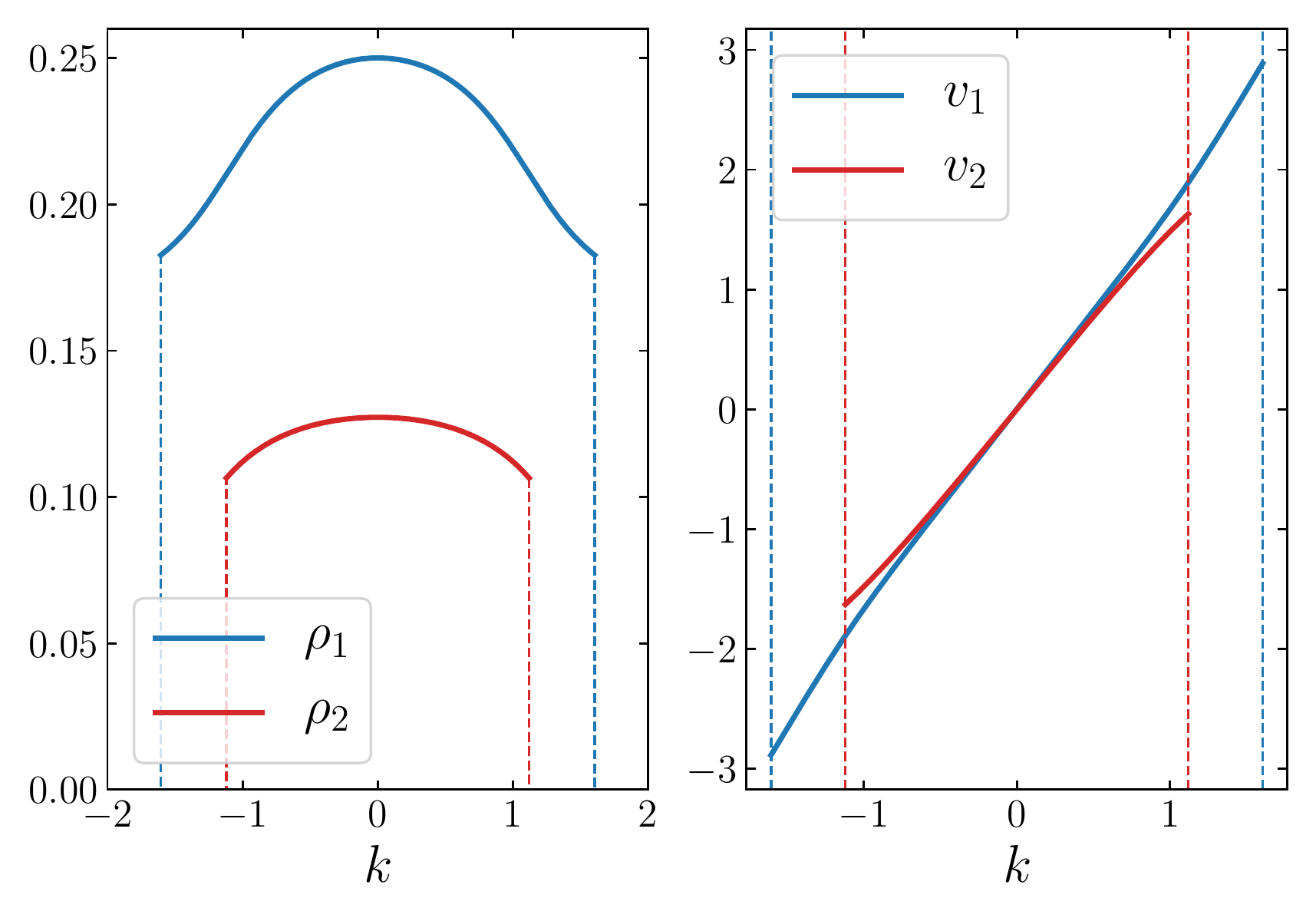} \;
	\includegraphics[width=0.45\textwidth]{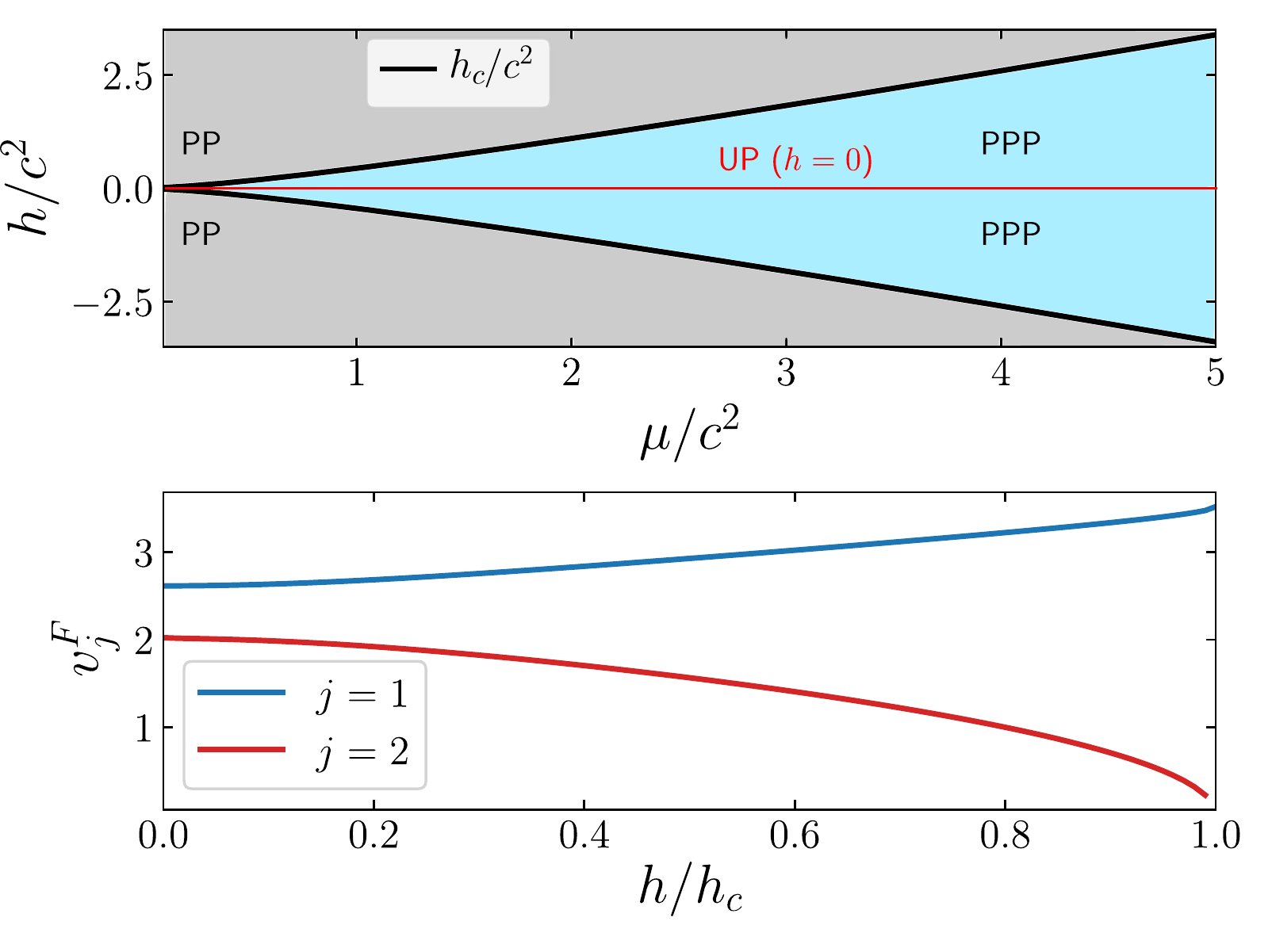}
	\caption{Panels (a),(b). Rapidity distribution functions (a) and effective velocities  (b) of the Yang-Gaudin model at zero temperature. The curves are obtained from the numerical solution of Eqs.~\eqref{eq:zero_BGT} (panel a) and Eqs.~\eqref{eq:zero_velocities_appendix} (panel b). The dashed vertical axes mark the position of the Fermi points $Q_{1,2}$. We have set $c=1$, $\mu=2$ and $h=0.5$. Panel (c)-top: Phase diagram of the Yang-Gaudin model at zero-temperature:  PP (gray regions), PPP (light-blue regions), and UP (red line). The phase boundary (black line) has been obtained by numerically solving Eq.~\eqref{eq:phase_boundary}. Panel (c)-bottom: Fermi velocities of the Yang-Gaudin model at zero temperature as a function of $h/h_c$. In the figure, we have set $\mu=2$ and $c=1$.}
	\label{fig:plot_zero_temperature}
\end{figure*}

As a key simplification, as $T\to 0$, one has $\eta_{2,n}(\lambda)\to\infty$ for $n\geq 2$, and the contribution of higher strings become negligible, i.e., $n_{2,n}(\lambda)\to 0$. Explicitly, setting $\eta_{1}(k)=e^{\epsilon_1(k)/T}$, and $\eta_{2,1}(k)=e^{\epsilon_{2}(k)/T}$, the limit $T\to 0$ of Eq.~\eqref{eq:TBA} yields
\begin{subequations}
\label{eq:zero_energy_TBA}
\begin{align}
\epsilon_1(k)&=k^2-\mu-h +[ \phi_1\ast \epsilon_{2}]^{Q_2}_{-Q_2}(k)\,, \quad k\in [-Q_1,Q_1]\,,\\
\epsilon_{2}(k)&=2h + [\phi_1 \ast \epsilon_1]^{Q_1}_{-Q_1}(k) -[ \phi_2  \ast \epsilon_{2}]^{Q_2}_{-Q_2}(k)\,,
\end{align}
\end{subequations}
with  $k\in [-Q_2,Q_2]$. Here, we introduced the notation
\be
[g_1\ast g_2]^A_{B}(k) = \int_{A}^{B} \dd k' \ g_1(k-k') g_2(k')\,,
\ee
while $Q_1$, $Q_2$ are cutoffs in the rapidity space known as Fermi points. They are determined self-consistently by the equation $\epsilon_{j}(Q_j)=0$. Note that the relation between the distribution of rapidities and holes becomes trivial in this limit. Namely, setting $n_{2}(\lambda)=n_{2,1}(\lambda)$, the Fermi factors acquire a step like form
\be\label{fermi-factor-zeroT}
n_j(q)=\left\{\begin{array}{ll}1 & \text { if } q\in[-Q_j,Q_j] \,, \\ 0 & \text { otherwise. }
\end{array}\right.
\ee
As a consequence, Eqs.~\eqref{eq:_appendix_BGT} and~\eqref{eq:velocities_TBA} simplify, and in the limit $T\to 0$ we get
\begin{subequations}
\label{eq:zero_BGT}
\begin{align}\label{eq:zero_BGT1}
\rho_1(k)&=\frac{1}{2\pi}+ [\phi_1\ast\rho_{2}]_{-Q_2}^{Q_2}(k)\,,\\
\rho_{2}(k)&=[\phi_1\ast\rho_1]_{-Q_1}^{Q_1}(k) - [\phi_2\ast\rho_{2}]_{-Q_2}^{Q_2}(k)\,,
\end{align}
\end{subequations}
and
\begin{subequations}
\label{eq:zero_velocities_appendix}
\begin{align}
v_1(k) \rho_1(k)&= \frac{k}{\pi} + [\phi_1\ast\rho_{2}v_2]_{-Q_2}^{Q_2}(k)\,,\\
v_{2}(k) \rho_{2}(k)&=[\phi_1\ast\rho_1 v_1]_{-Q_1}^{Q_1}(k) - [\phi_2\ast\rho_{2} v_{2}]_{-Q_2}^{Q_2}(k)\,.
\end{align}
\end{subequations}
We report an example of the numerical solution to these equations in Fig.~\ref{fig:plot_zero_temperature}.

Equation~\eqref{eq:zero_energy_TBA} gives straightforward access to the ground-state phase diagram of the model, cf. Fig.~\ref{fig:plot_zero_temperature}. It is characterized by a critical line of the magnetic field, defined by the equation~\cite{batchelor2010exactly},
\begin{align}\label{eq:phase_boundary} 
h_c +& \frac{1}{\pi} \left[\frac{c}{2}\sqrt{\mu + h_c} \right.\nonumber\\
-&\left.(\mu +h_c+ \frac{c^2}{4}) \tan^{-1}\left(\frac{\sqrt{\mu + h_c}}{c/2}\right)\right]=0\,.
\end{align}
More precisely,  on varying the magnetic field $h$, one finds three distinct phases:
\begin{itemize}
	\item[-]{\it Polarized phase} (PP) for $h\geq h_c$: Ferromagnetic ground state without spin waves ($Q_2=0$). The first species behaves as a gas of spinless noninteracting particles with Fermi point given by $Q_1=\sqrt{\mu+h}$.
	\item[-]{\it Partially polarized phase} (PPP) for $h_c<h<0$: The ground state is paramagnetic, both species are present. The Fermi points $Q_{1,2}$ are extracted from Eq.~\eqref{eq:zero_energy_TBA}, using the self-consistent condition $\epsilon_{j}(Q_j)=0$.
	\item[-]{\it Unpolarized phase} (UP) at $h=0$. In this case, both species are present, and the Fermi rapidity of the second one goes to infinity, $Q_{2}\to \infty$.
\end{itemize}

It is interesting to consider the behavior of the Fermi velocities $v^F_j=v_j(Q_j)$ in the limit $h\to 0$, which we plot in Fig.~\ref{fig:plot_zero_temperature}. We see that, although $Q_2\to\infty$ as $h\to 0$, $v_{2}^F$ remains finite. As we increase the magnetic field, the velocity of the first (second) species increases (decreases), and for $h=h_c$ the spin excitations become infinitely slow. Importantly,  the two velocities are always separated, which is a crucial feature to observe SCS.

\section{The GHD equations and the numerical solution}
\label{sec:GHD_equations}

GHD is a hydrodynamic theory which is based on a description of local fluid cells in terms of rapidity distribution functions~\cite{bertini2016transport,castro-alvaredo2016emergent}. One of its main results is the corresponding set of continuity equations, which in our case read~\cite{mestyan2019spin}
\begin{subequations}\label{eq:appendix_GHD}
	\begin{align}
		\label{eq:appendix_GHD1}
		\de_t \rho_1(k) + \de_x [v_1(k)\rho_1(k)] =\left(\partial_{x} V\right) \partial_{k} \rho_1(k)\,,\\
		\label{eq:appendix_GHD2}
		\de_t \rho_{2,n}(\lambda) + \de_x [v_{2,n}(\lambda) \rho_{2,n}(\lambda)] =\left(\partial_{x} V\right) \partial_{\lambda} \rho_{2,n}(\lambda)\,.
	\end{align}
\end{subequations}
In order to obtain a numerical solution to these equations, it is more convenient to consider the equivalent formulation in terms of the Fermi factors~\eqref{eq:appendix_fermi_factors}. Following standard derivations~\cite{bertini2016transport, castro-alvaredo2016emergent}, one obtains
\begin{subequations}\label{eq:ghd}
\begin{align}
\left(\de_t  + v_1 \de_x - (\de_x V) \de_k\right) n_1=0\,,\label{eq:ghd1}\\
\left(\de_t  + v_{2,n} \de_x- (\de_x V) \de_\lambda \right)n_{2,n} =0\,.\label{eq:ghd2}
\end{align}
\end{subequations}
Now Eq.~\eqref{eq:ghd} is formally solved for $\bm n=\{n_1,n_{2,n}\}$ using the method of characteristics, yielding
\be
\bm n(t,x,k)= \bm n(0, \tilde{x}(t), \tilde{k}(t)),
\ee
with
\begin{subequations}
\begin{align}
\tilde{x}(t)= x -\int_0^t \dd s \ \bm{v}(s,\tilde{x}(s),\tilde{k}(s)),\\
\tilde{k}(t)=k-\int_0^t \dd s \ \bm{a}(\tilde{x}(s)).
\end{align}\end{subequations}
Here, $\bm{v}=\{ v_1, v_{2,n}\}$ and $\bm{a}=\{ -\de_x V, -\de_x V\}$ is the effective acceleration. This is the starting point for an efficient numerical solution to the GHD equations~\cite{frederik2020introducing}. In our work, we have implemented this approach to obtain the numerical data presented in Fig.~\ref{fig:profile_zero_temperature}. In particular, we have followed the steps outlined in Ref.~\cite{bastianello2019generalized}. Since our implementation is standard, we refer the reader to the latter work for more details. \\

We note that, for finite temperature, the numerical solution to the GHD equations is more demanding compared with the case of the Lieb-Liniger model~\cite{Caux2019QNC}. This is particularly true for $h=0$, where one needs to retain a large number of strings in order to obtain accurate numerical results~\cite{takahashi2005thermodynamics,mestyan2019spin}. In practice, we have always solved the GHD equations keeping a finite number of them, and checked that, within the desired numerical accuracy, the final result did not change by increasing the number of strings considered. At $h=0$ we found that the magnetization exhibited the largest numerical inaccuracy compared with other observables. For the profiles shown in this work, we have computed the magnetization up to an absolute inaccuracy $\delta \sim 10^{-3}$, while other quantities are always determined to higher precision. These numerical errors can be made smaller by increasing the number of strings and of discretization points, although this requires longer computational times.\\

At zero temperature, the numerical solution of the GHD equations greatly simplifies thanks to the zero-entropy condition of the fluid cells \cite{doyon2017anote}. Thus, for a complete description of the GHD evolution, it is sufficient to consider the points $(x, \theta_{\alpha}^{(j)})$ in the rapidity-position plane, whose dynamics is given by \cite{doyon2017large}
\be\label{eq:phase-space}
\frac{\dd}{\dd t}\begin{pmatrix} x \\[4pt] \theta_\alpha^{(j)} \end{pmatrix} = \begin{pmatrix} v_j(t,x,\theta^{(j)}_{\alpha}) \\[4pt] - \de_x V \end{pmatrix} .
\ee
From this equation, we have implemented a zero-entropy GHD algorithm following the strategy of Ref.~\cite{doyon2017large}. The numerical data in Figs.~\ref{fig:contour_evolution} and \ref{fig:profile_evolution} are obtained using this procedure.

\section{Perturbative corrections to SCS}
\label{sec:perturbative_scs}

In this appendix, we discuss in detail the corrections to the SCS, in the context of zero-entropy GHD.

We begin by analyzing the zero-entropy BE, in the case where both species display a single Fermi sea, and the Fermi rapidities of the second one are at infinity, namely,
\begin{subequations}\label{eq:one}
	\begin{align}
		\rho_1(k)&=\frac{1}{2\pi}+ [\phi_1\ast\rho_{2}]_{-\infty}^{+\infty}(k),\\
		\rho_{2,1}(\lambda)&=[\phi_1\ast\rho_1]_{A_-}^{A_+}(\lambda) - [\phi_2\ast\rho_{2,1}]_{-\infty}^{+\infty}(\lambda)\,,\label{eq:aux_2}
	\end{align}
\end{subequations}
where we take $A_{\pm}$ to be arbitrary. Let us first show that $\rho_{2,1}(\lambda)$ vanishes exponentially in $\lambda$, as $\lambda\to\infty$. To do this, we expand
\begin{align}
[\phi_1\ast\rho_1]_{A_-}^{A_+}(\lambda)=&\frac{1}{\pi}\frac{2c}{c^2 + 4\lambda^2}\int_{A_-}^{A_{+}}\dd k \rho_1(k)+O(1/\lambda^3)\nonumber\\
=&a_1(\lambda)\varrho+O(1/\lambda^3)\,.
\end{align}
Plugging this into the rhs of Eq.~\eqref{eq:aux_2}, and keeping only the leading term, we obtain that $\rho_{2,1}(\lambda)$ has the same large-$\lambda$ behavior of the function $\tilde{\rho}_{2,1}(\lambda)$, which satisfies
\be
\tilde{\rho}_{2,1}(\lambda)=a_{1}(\lambda)\varrho-\int_{-\infty}^{\infty}\dd \mu\, a_{2}(\lambda-\mu)\tilde{\rho}_{2,1}(\lambda)\,.
\ee
This equation is readily solved in Fourier space, yielding $\tilde{\rho}_{2,1}(\lambda)=(\varrho/4) \operatorname{sech}(\pi \lambda /c)$, from which we obtain the leading behavior 
\be\label{eq:estimate_1}
\rho_{2,1}(\lambda)\propto e^{-\pi |\lambda|/c}\,.
\ee

Now, let us consider the large-time profiles of charge and spin displayed in Fig.~\ref{fig:profile_evolution}. We first focus on the main peak of the charge at large $x>0$, corresponding to a zero-entropy state characterized by the Fermi rapidities $A_\pm$, and $B_\pm$,  with $|B_-|>|B_+|$. We wish to show that the effects of the charge pulse propagation on the magnetization are exponentially small in $|B_{+}|$, proving SCS as discussed in the main text. In this case, the equations to solve are
 \begin{subequations}\label{eq:two}
	\begin{align}
		\rho^{(B_{\pm})}_1(k)&=\frac{1}{2\pi}+ [\phi_1\ast\rho^{(B_{\pm})}_{2}]_{B_-}^{B_+}(k)\,,\\
		\rho^{(B_{\pm})}_{2,1}(\lambda)&=[\phi_1\ast\rho^{(B_{\pm})}_1]_{A_-}^{A_+}(\lambda) - [\phi_2\ast\rho^{(B_{\pm})}_{2,1}]_{B_-}^{B_+}(\lambda)\,.\label{eq:aux_3}
	\end{align}
\end{subequations}
We need to show that the magnetization is vanishing, up to terms that are exponentially small in $|B_+|$, namely
\be\label{eq:to_prove}
\int_{B_{-}}^{B_+}\dd k\, \rho^{(B_{\pm})}_{2,1}(k)-\frac{1}{2}\int_{A_{-}}^{A_+}\dd k\, \rho^{(B_{\pm})}_{1}(k)=O(e^{-\pi |B_{+}|/c})\,.
\ee
Defining
\begin{align}
\delta \rho_1(k)&=\rho^{(B_{\pm})}_1(k)- \rho^{(\pm \infty)}_1(k)\,,\\
\delta \rho_{2,1}(k)&=\rho^{(B_{\pm})}_{2,1}(k)- \rho^{(\pm \infty)}_{2,1}(k)\,,
\end{align}
and subtracting Eq.~\eqref{eq:one} from Eq.~\eqref{eq:two}, we obtain
 \begin{subequations}\label{eq:difference_equations}
	\begin{align}
		\delta \rho^{(B_{\pm})}_1(k)&=g_1(k)+ [\phi_1\ast \delta\rho^{(B_{\pm})}_{2}]_{B_-}^{B_+}(k)\,,\label{eq:difference_equations_1}\\
		\delta\rho^{(B_{\pm})}_{2,1}(\lambda)&=-g_2(\lambda)+[\phi_1\ast \delta\rho^{(B_{\pm})}_1]_{A_-}^{A_+}(\lambda) \nonumber\\
		-& [\phi_{2}\ast \delta \rho^{(B_{\pm})}_{2,1}]_{B_-}^{B_+}(\lambda)\,,
	\end{align}
\end{subequations}
where
\begin{align}
g_1(k)=\int_{\mathbb{R}\setminus [B_{-},B_+]} \dd{\lambda}\,\phi_1(k-\lambda)\rho_{2,1}^{(\pm \infty)}(\lambda)\sim e^{-\pi |B_{+}|/c}\,,\\
g_2(k)=\int_{\mathbb{R}\setminus [B_{-},B_+]} \dd{\lambda}\,\phi_2(k-\lambda)\rho_{2,1}^{(\pm \infty)}(\lambda)\sim e^{-\pi |B_{+}|/c}\,,
\end{align}
where we used Eq.~\eqref{eq:estimate_1}. Thus, $\delta \rho^{(B_{\pm})}_1(k)$ and $ \rho^{(B_{\pm})}_{2,1}(k)$ satisfy the same integral equations as $\rho^{(B_{\pm})}_1(k)$ and $\delta \rho^{(B_{\pm})}_{2,1}(k)$, but the driving terms are exponentially vanishing in $|B_{+}|$.  Therefore, $\delta \rho_1(k)$,  $\delta \rho_{2,1}(k)=O(e^{-\pi|B_+|/c})$, and   
\begin{align}
\int_{A_{-}}^{A_+}\dd k\, \rho^{(B_{\pm})}_{1}(k)&=\int_{A_{-}}^{A_+}\dd k\, \rho^{(\pm \infty)}_{1}(k)+O(e^{-\pi|B_+|/c})\,,\label{eq:to_test1}\\
\int_{B_{-}}^{B_+}\dd k\, \rho^{(B_{\pm})}_{2,1}(k)&=\int_{B_{-}}^{B_+}\dd k\, \rho^{(\pm \infty)}_{2,1}(k)+O(e^{-\pi|B_+|/c})\nonumber\\
&=\int_{-\infty}^{\infty}\dd k\, \rho^{(\pm \infty)}_{2,1}(k)+O(e^{-\pi|B_+|/c})\,,\label{eq:to_test2}
\end{align}
where the last equality in the second line follows from Eq.~\eqref{eq:estimate_1}. Multiplying now Eq.~\eqref{eq:to_test1} by $1/2$ and subtracting Eq.~\eqref{eq:to_test2}, we get
\begin{widetext}
\be\label{eq:almost_there}
\int_{B_{-}}^{B_+}\dd k\, \rho^{(B_{\pm})}_{2,1}(k)-\frac{1}{2}\int_{A_{-}}^{A_+}\dd k\, \rho^{(B_{\pm})}_{1}(k)=\int_{-\infty}^{\infty}\dd k\, \rho^{(\pm \infty)}_{2,1}(k)-\frac{1}{2}\int_{A_{-}}^{A_+}\dd k\, \rho^{(\pm \infty)}_{1}(k)+O(e^{-\pi |B_{+}|/c})\,.
\ee
As a last step, integrating both sides of Eq.~\eqref{eq:aux_2} over $\mathbb{R}$ and using $\int_{-\infty}^{+\infty}\dd \lambda\, a_{n}(\lambda)=1$, we obtain
\be\label{eq:temp_2}
\int_{-\infty}^{\infty}\dd k\, \rho^{(\pm \infty)}_{2,1}(k)=\int_{A_{-}}^{A_+}\dd k\, \rho^{(\pm \infty)}_{1}(k)-\int_{-\infty}^{\infty}\dd k\, \rho^{(\pm \infty)}_{2,1}(k)\,.
\ee
Combining~\eqref{eq:almost_there} with~\eqref{eq:temp_2} we finally obtain~\eqref{eq:to_prove}. 

\begin{figure*}
	\centering
	(a) \hspace{4cm} (b) \hspace{4cm} (c)\\
	\includegraphics[width=0.8\textwidth]{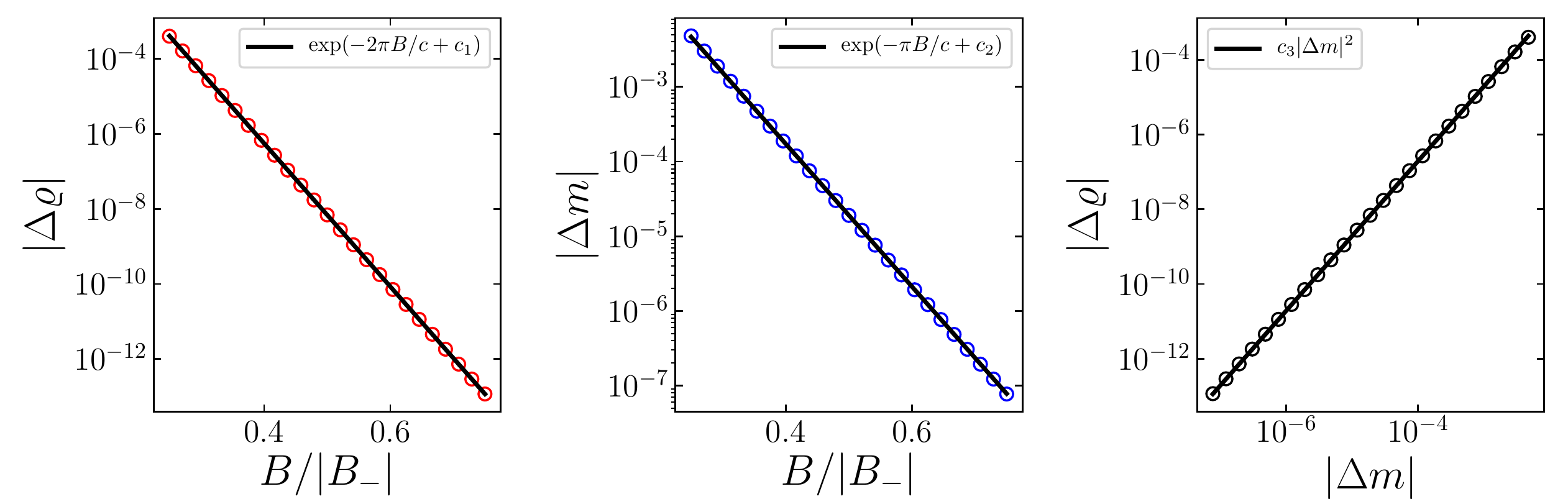} 
	\caption{(a) Log-plot showing $|\Delta \rho|$ as a function of $B/|B_-|$ (circles). The line corresponds to the fitting function $y(B)=\exp(-2\pi B/c +c_1)$ with $c_1\simeq 3.2$ extracted from our data. (b) Log-plot showing $|\Delta m|$ as a function of $B/|B_-|$ (circles). The line corresponds to $y(B)=\exp(-\pi B/c +c_2)$ with $c_2\simeq 0.1$. (c) Log-log-plot showing $|\Delta \varrho|$ as a function of $|\Delta m|$ (circles). The line corresponds to $y(|\Delta m|)= c_3|\Delta m|^2$ with $c_3\simeq 2.9$. The figures are obtained with $c=1$, $\mu=1.5$, and $h=0$ and setting the UV cutoff $B_-=-10$.}
	\label{fig:log-log_plots}
\end{figure*}

Finally, let us consider the main peak of the magnetization profile at large times and $x>0$, cf. Fig.~\ref{fig:profile_evolution}. 
This corresponds to a zero-entropy state where the two species have Fermi rapidities $\pm k_0$ and $B_\pm$, respectively. Since $|B_-|\gg |B_{+}|$,  neglecting subleading contributions we can assume for simplicity $B_-=-\infty$. Setting $B_+=B$, we define

\begin{align}
\Delta \varrho&= \int_{-k_0}^{k_0}\dd k \rho^{(B)}_{1}(k)-\int_{-k_0}^{k_0}\dd k \rho^{(\pm \infty)}_{1}(k)\,,\\
\Delta m&= \left[\int_{-\infty}^{B}\dd \lambda \rho^{(B)}_{2,1}(\lambda)-\frac{1}{2}\int_{-k_0}^{k_0}\dd k \rho^{(B)}_{1}(k)\right]-\left[\int_{-\infty}^{+\infty}\dd \lambda \rho^{(\pm \infty)}_{2,1}(\lambda)-\frac{1}{2}\int_{-k_0}^{k_0}\dd k \rho^{(\pm \infty)}_{1}(k)\right]\,,
\end{align}
\end{widetext}
where we denoted by $\rho_{1}^{(B)}(k)$ and $\rho_{2,1}^{(B)}(\lambda)$ the solution to Eq.~\eqref{eq:two} with $B_-=-\infty$, $B_+=B$ and $A_\pm=\pm k_0$. Clearly, we have
\be
\lim_{B\to\infty}|\Delta \varrho|=\lim_{B\to\infty}|\Delta m|= 0\,.
\ee 
In principle, it is possible to obtain analytically the exact leading behavior of $|\Delta \varrho|$ and $|\Delta m|$ for large $B$ from Eqs.~\eqref{eq:difference_equations}. However, we found that this is nontrivial. One of the reasons is that the two terms in the rhs of Eq.~\eqref{eq:difference_equations_1} have opposite signs, and subtle cancellations happen at the leading orders in $B$. Because of this, we found it more convenient to perform a numerical analysis, from which the leading behavior emerges clearly. Indeed, Eq.~\eqref{eq:two} can be solved to very high precision using the Gaussian quadrature method~\cite{vetterling2002numerical}, which allows us to obtain very reliable data even for small values of $|\Delta m|$. An example of our finding in given in Fig.~\ref{fig:log-log_plots}, from which we clearly see that both $|\Delta m|$ and $|\Delta \varrho|$ decay exponentially in $B$.  From inspection of Eq.~\eqref{eq:difference_equations}, we expect $|\Delta m|,|\Delta \varrho|= O( e^{-\pi B/c})$. In fact, we find that the exact behavior is  given by 
\be\label{eq:behavior}
|\Delta m|\sim e^{-\pi B/c}\,,\qquad |\Delta \varrho|\sim e^{-2\pi B/c}\,.
\ee 
Namely, for the charge $|\Delta\varrho|$ nontrivial cancellations happen and the leading behavior is found to be $e^{-2\pi B/c}$. We confirmed this to high precision with a linear fit in logarithmic scales, where the only free parameter is the value of the additive constant. Finally, as an immediate consequence of Eq.~\eqref{eq:behavior}, we have that for large values of $B$ 
\be\label{eq:final_result}
|\Delta \varrho|\sim |\Delta m|^2\,,
\ee
as announced in the main text.  We have have also tested directly Eq.~\eqref{eq:final_result} finding excellent agreement, cf. Fig.~\ref{fig:log-log_plots}.

\bibliography{./bibliography}

\end{document}